Draft

# Modeling and simulation of a beam emission spectroscopy diagnostic for the ITER prototype neutral beam injector [a]

M. Barbisan,[1,b] B.Zaniol,[1] and R. Pasqualotto[1]

[1]*Consorzio RFX, C.so Stati Uniti 4, 35127 Padova, Italy*

A test facility for the development of the Neutral Beam Injection system for ITER is under construction at Consorzio RFX. It will host two experiments: SPIDER, a 100 keV $H^-/D^-$ ion RF source, and MITICA, a prototype of the full performance ITER injector (1 MV, 17 MW beam). A set of diagnostics will monitor the operation and allow to optimize the performance of the two prototypes. In particular, Beam Emission Spectroscopy will measure the uniformity and the divergence of the fast particles beam exiting the ion source and travelling through the beam line components. This type of measurement is based on the collection of the $H\alpha/D\alpha$ emission resulting from the interaction of the energetic particles with the background gas. A numerical model has been developed to simulate the spectrum of the collected emissions in order to design this diagnostic and to study its performance. The paper describes the model at the base of the simulations and presents the modeled $H_\alpha$ spectra in the case of MITICA experiment.
[DOI]

## I. INTRODUCTION

Neutral Beam Injectors (NBI) are a component of the ITER experiment which will be crucial to take the reactor to its ultimate performance. Each one of these devices must produce a beam of deuterium particles of 1 MeV energy for an overall power of 17 MW. Presently the most efficient way to get this consists in producing $H^-/D^-$ ions with a cesiated RF source and in accelerating them up to the nominal energy through a system of grids. Most of the ions are then converted into neutrals in the so-called neutralizer, by means of a charge-exchange process with $H_2/D_2$ gas. At last, all the remaining ions are eliminated in the electrostatic Residual Ion Dump (RID) before they can enter the reactor. To face the technical and scientific complexity of this system, a full performance prototype of the injectors, MITICA[1], is under construction at Consorzio RFX. In addition, a 100 keV RF negative ion source, SPIDER[2], will be built prior to MITICA to support its design and its future operation. The beam of negative ions produced inside SPIDER and MITICA will be composed by 1280 smaller beamlets, as many as the holes of the acceleration grids. The direction and the divergence of the beamlets will have to be continuously monitored because they determine the dimension and the uniformity of the beam. A too large or badly-shaped beam could indeed hit and damage parts of the injector, while non-uniformities would compromise the effectiveness of the beam. This information will be retrieved mainly by three diagnostics[3]: the calorimeter which stops the beam when inserted, visible tomography and beam emission spectroscopy (BES). This last consists in the spectral analysis of the radiation produced by the interaction of the beam with the background gas. To support the design of this diagnostic a numerical model has been developed, from which it is possible to simulate the spectra acquired by the diagnostic itself. This paper will focus on BES diagnostic for the MITICA experiment. After a brief recall of the principles of beam emission spectroscopy, it will explain how the diagnostic parameters have been optimized to minimize the measurement errors. Subsequently, it will be shown how the beam and the optics of the diagnostic have been simulated. The resulting spectra and their analyses will be eventually shown and discussed.

## II. PRINCIPLE OF OPERATION OF BES

Beam emission spectroscopy consists in observing the $H_\alpha/D_\alpha$ radiation produced by the collisions of the beam particles with the background gas. The collisions can lead to the dissociation of the gas molecules and the excitation to the n=3 level of their atoms; the following de-excitation to the n=2 level of the atom generates a photon at the nominal wavelength $\lambda_0$ (656.28 nm for hydrogen, 656.10 nm for deuterium). The beam particles can also be excited by gas collisions and the wavelength $\lambda'$ of the resulting radiation is Doppler shifted according to the following equation:

$$\lambda' = \lambda_0 \frac{1-\beta\cos\alpha}{\sqrt{1-\beta^2}}, \quad \beta = \frac{v}{c} \qquad (1)$$

where v is the speed of the beam particle, c is the speed of light and α is the angle between the direction of the emitting particle and the direction of propagation of the produced photon. In MITICA[1] the beam particles will have a maximum energy of 870 keV in the case of hydrogen and 1 MeV in the case of deuterium, corresponding to values of β of $4.30 \times 10^{-2}$ and $3.26 \times 10^{-2}$, respectively. The average value of α is given by the direction of observation, since the light is collected by a number of telescopes along as many lines of sight (LOSs). The telescopes are collimating lenses coupled to optical fibers which, at the other end, are piled along the entrance slit of two grating spectrometers; these are coupled to 2D CCD cameras to simultaneously record the spectra of all the LOSs. The simplest information measurable from a spectrum is the direction of the beamlets in terms of α. This angle can indeed be calculated from equation 1, using β which is always known from the voltage of the acceleration grids, and employing the wavelength separation between the Doppler shifted peak and the unshifted one. The divergence ε of the beamlets, instead, can be measured from the spectral width $\Delta\lambda$ of the shifted $H_\alpha$ component. This line, indeed, is broadened[4] by means of the Doppler effect by quantities which affect β and α in equation 1: the voltage ripple υ of the grids, the already mentioned ε and the angle ω of collection of the photons due to the finite dimension of the lens of the telescopes. These contributions must be quadratically summed[4] to the intrinsic width $\Delta\lambda_N$=13.2 pm of the $H_\alpha$ line and to the broadening $\Delta\lambda_I$ given by the instrumental function of the spectrometer:

$$\Delta\lambda = \left\{ \Delta\lambda_I^2 + \Delta\lambda_N^2 + \left(\frac{\lambda_0}{\sqrt{1-\beta^2}}\beta\sin\alpha\right)^2(\omega^2+\varepsilon^2) + \left[\frac{e\lambda_0}{mc^2\beta}(\beta-\cos\alpha)\right]^2 \upsilon^2 \right\}^{1/2} \quad (2)$$

where m and e are the mass and the charge of the electron, respectively.

## III. DESIGN OF THE DIAGNOSTIC

The BES diagnostic for MITICA, as in the case of SPIDER[5], must be able to measure the uniformity and the divergence of the beam with an error not greater than 10%. The measurement of ε is the main target which has driven the diagnostic design. The divergence can be estimated from $\Delta\lambda$ only if all the other parameters contributing to the broadening (Eq. 2) are known. Their estimates, together with their experimental accuracies ($\Delta\lambda$ included), determine the measurement error of ε [4]. The basic information needed to estimate this error is listed in Table 1; in particular, it was assumed that the CCD and the spectrometer will be of the same model of those adopted in the BES diagnostic for SPIDER[4,5]. Assuming a typical value of 5 mrad for ε, the error $\sigma_\varepsilon$ of the divergence is practically set by α and ω; Figure 1a shows this dependence in the case of deuterium, since it gives higher values of $\sigma_\varepsilon$ compared to hydrogen. The best compromise was found adopting 85° observation angle and about 2.2 mrad optical aperture. A higher value of α wouldn't give a significant improvement to the accuracy of ε and would reduce the capability to distinguish the contributions of beamlets with slightly different values of α. This issue is important because in MITICA the beamlets won't be parallel to each other (see chap. VI). In the opposite way, a lower α would give an higher Doppler shift and would therefore allow to better discriminate the beamlets direction; this, however, at the cost of increasing the relative error of ε. Regarding ω, instead, a lower optical aperture would affect the intensity of the spectrum and then the achievable time resolution, whereas a higher value of ω would increase $\sigma_\varepsilon/\varepsilon$. In Figure 1b it is possible to see how the relative error of the divergence varies as function of α and the divergence itself. For α=85°, $\sigma_\varepsilon/\varepsilon$ should be 10%, 4% and 2% respectively for the minimum, typical an maximum expected values of ε. The plot shows also that the error would strongly increase (especially at low divergence) if α were reduced with respect to the chosen value.

**Table 1:** Fixed parameters relevant to the estimate of the divergence accuracy.

| Quantity | Value |
|---|---|
| Divergence (e-folding) | 3÷7 mrad ±10% max. (5 mrad typical) |
| Typical measurement error of $\Delta\lambda$ with a SNR around 50 | 0.2 pixel |
| Plate factor of the spectrometer | $(6.67\pm0.01)\cdot10^{-3}$ nm/pixel |
| Default entrance slit width of the spectrometer | 100 µm |
| CCD pixel size | 13 µm |
| Typical beamlets distance from the lens | 1.1÷2.2 m |
| Variability of ω due to width and different position of beamlets | 8 % |
| Acceleration voltage | 870 kV (H), 1 MV (D) ±1% |
| Voltage ripple | < 5.0 ± 0.5 % |
| Typical measurement error of α from BES spectra | ~2 mrad |

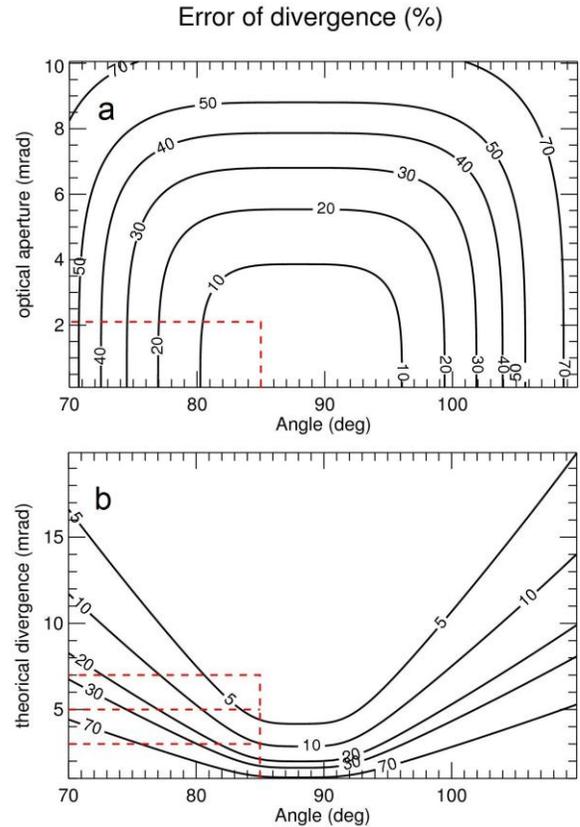

**Figure 1:** Measurement accuracy of ε expected for D operation, as function of the angles α and ω in (a) and of α and ε itself in (b). The selected values of α, ω and ε are indicated by dashed lines.

## IV. MODELING OF THE BEAM

Once the design parameters are set, it is possible to simulate the spectra obtained by the BES diagnostic. The first step consists in numerically reproducing the properties of the beamlets. Firstly, according to ITER prescriptions, the beam axis will be aimed 49 mrad downwards. In the case of ITER injectors and MITICA, the directions of the single beamlets are not parallel to the beam axis but follow a complex geometry to focus the beamlets at the torus entrance and to improve the overall uniformity of the beam in the plasma. If projected on a vertical plane, the beamlets form angles of 4 prefixed values with the axis of the beam, one for each horizontal sector of the grids, whereas on an horizontal plane there can be 20 different angles, as the 5 beamlets of each column focus at the exit of the RID[6]. The BES diagnostic, as in SPIDER[5], will have LOSs vertically and horizontally oriented, therefore the respective spectra will have up to 4 and 20 Doppler shifted components, respectively. At the exit of the acceleration grids each beamlet is modeled with 3 separate components:
- negative ions at full energy and low divergence (expected between 3 mrad and 7 mrad e-folding);
- ions which have been neutralized inside the acceleration stage (the so-called stripping losses) and have therefore lower energy and higher divergence;
- a "halo" of badly focused ions ($\varepsilon \leq$ 30 mrad e-folding).

The overall fraction of stripping losses, as well as their energetic distribution, is calculated by the program AVOCADO[7,8]. The halo, instead, is neglected at the moment. The target extracted current density J is 355 A/m$^2$ for hydrogen and 285 A/m$^2$ for deuterium. The density $n_b$ of each component of the beamlets, assuming that its transversal profile is Gaussian, can be calculated at any point in space:

$$n_b(\delta,l) = \frac{Jr_G^2}{2\sigma_b^2(l)v} e^{-\frac{\delta^2}{\sigma_b^2(l)}}, \quad \sigma_b(l) = \sigma_0 + l\tan\left(\frac{\varepsilon}{\sqrt{2}}\right) \quad (3)$$

where $r_G$=7 mm is the radius of the holes of the last acceleration grid, $\delta$ is the distance from the beamlet axis, $\sigma_0$=4 mm and $\sigma_b$ are the Gaussian widths of the beamlet respectively at the exit of the last grid and at a distance l along the beamlet axis. Along their trajectory the beamlets interact with the background gas and the ionic composition of their components varies: the H$^-$/D$^-$ ions can become neutrals or H$^+$/D$^+$, the neutrals can be positively ionized and the H$^+$/D$^+$ can be neutralized. The composition of the ions is important because different ions interact with the gas and produce H$_\alpha$/D$_\alpha$ radiation with different cross sections[9]. This is particularly important in the measurements of beam uniformity, where H$_\alpha$/D$_\alpha$ emissions along different LOSs are compared.

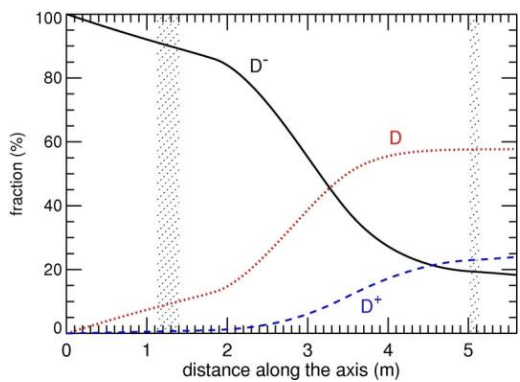

**Figure 2:** Ionic composition (D$^-$: solid line, D$^0$: dotted line, D$^+$: dashed line) of an hypothetical deuterium beamlet placed on the axis of the beam. The dashed areas indicate the position of the LOSs. Beyond 5.4 m all the ions are captured by the RID.

The evolution of the 3 ionic species can be calculated solving the set of differential equations which describe the detailed balance of the processes of ionization/neutralization. The solutions depend on the cross sections of the reactions[9] and on the density $n_{gas}$ of the background gas (calculated by AVOCADO[7,8]), which varies mainly along the axis of the vacuum vessel. An example solution is shown in Figure 2, for deuterium ions and using the axis of the beam as trajectory. The calculation must however be performed for each beamlet, because the profile of $n_{gas}$ must be projected on the specific trajectory of the beamlet itself.

## V. MODELING OF THE OPTIC SETUP

After having defined the position and the extension of the beamlets, it is necessary to determine which part of them is effectively observed by the LOSs of the diagnostic. The telescopes are composed by a lens which conveys the light to an optical fiber of 400 μm diameter, in turn connected to a spectrometer. The lens has a focal length of 150 mm and its clear aperture diameter is 8 mm. The fiber end is imaged at 3 m, a distance which is beyond any crossed beamlet. The volume of the LOSs is assumed to be a truncated cone, whose bases are the lens aperture and the image of the fiber head at 3 m. Each component of each beamlet is separately treated, and is considered only if its distance from the volume of the LOS is lower than $3\sigma_b$; for each selected component the following volume integral is calculated:

$$I = \int_{V_{LOS}} \frac{\Omega}{4\pi} n_{gas} n_i \, dV \quad (4)$$

where $V_{LOS}$ is the volume of the LOS, $n_i$ is the density of the i-th beamlet component (calculated as in equation 3) and $\Omega$ is the solid angle subtended by the lens of the telescope. Both $n_{gas}$, $n_i$ and $\Omega$ vary in the 3D space; in particular, $\Omega$ is only function of the distance from the lens and from the axis of the LOS itself. The calculation is performed considering the section p of the LOS at a given distance and the image q of the fiber head. Both areas are discretized according to polar coordinates. Let's consider a bunch of light rays traveling from a certain differential area of p to each of the elements of q: only some of the rays will fall in the area of the lens and will be therefore collected. The total solid angle with which the valid elements of q are subtended by the element of p gives $\Omega$ in that point of p. The resulting values are shown in Figure 3; the irregularities at high distance are not real but only a side effect of the discretization of the planes.

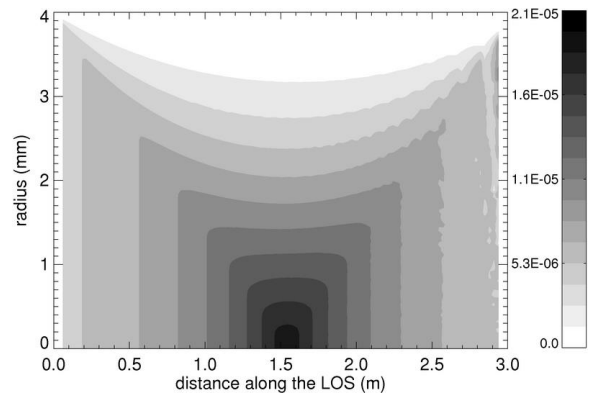

**Figure 3:** Solid angle (srad) of observation of the lens, as function of the distance from the lens and of the radial distance to the axis of the LOS.

## VI. RESULTS OF THE SIMULATIONS

Once the integral I has been calculated for all the components of the beamlets, the resulting values are multiplied by the speed of the particles, and by the cross sections[9] related to the production of $H_\alpha/D_\alpha$ photons. The energetic distribution of the stripping losses is taken into account, and each reaction is separately considered, distinguishing the type of ion involved and whether the excited particle belongs to the beam or to the gas. Each contribution is then added to the simulated spectrum as a Gaussian curve whose centroid and width are given respectively by equations 1 and 2; in particular, to calculate the width of the gaussians the values of ω are obtained by the values of Ω at the axis of the LOS. At last, the spectrum is scaled considering the losses of the optical instrumentation, the quantum efficiency of the CCD and its exposure time. The effects of the bias of the CCD and of the noise are added, too. Two typical spectra obtained in this way are shown in Figure 4.

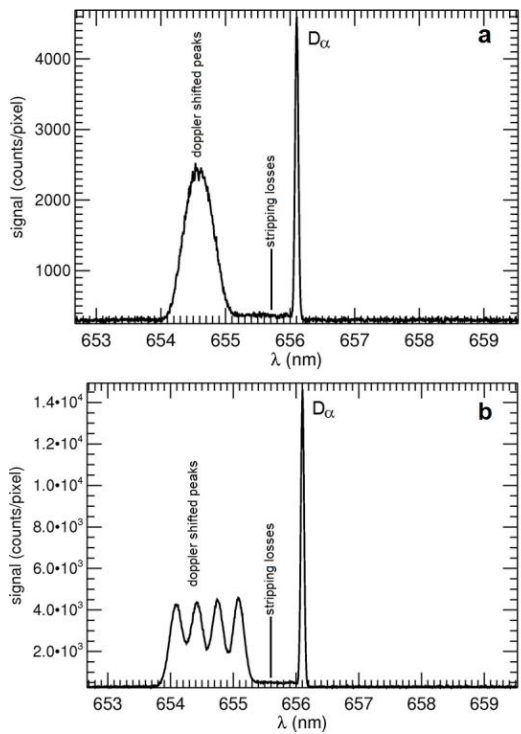

**Figure 4:** Typical spectra obtained by the simulations for an horizontal (a) and a vertical (b) LOS, assuming ε=5 mrad. The time exposure of the CCD was set to 150 ms.

As mentioned in chapter IV, the Doppler shift measured by an horizontal LOS can have 20 different values; the resulting peaks are however so close to each other that their sum corresponds to a single broader line (Figure 4a). To retrieve the average width (and therefore ε) of the single peaks it is necessary to fit the spectral line with the sum of 20 gaussians whose centroids are predetermined by the known values of α and β. The measurement accuracy of ε obtained with this method is plotted in Figure 5 (triangles) as function of the divergence itself. The relative error of ε is higher than expected (dashed line) in the desired interval of ε but still about 10%. This is due to the type of fitting function and to the contribution of the stripping losses (beam particles not accelerated at the maximum energy) which distorts the background level. With the vertical LOSs instead α has only 4 different values (one for each row of beamlet groups), corresponding to as many peaks which can be easily resolved in the spectrum (Figure 4b). In this situation the group of peaks can be fitted with the sum of 4 Gaussians, so that 4 different linewidths and therefore 4 values of ε can be obtained. The average of the relative errors of these 4 measures is shown in Figure 5 (diamonds) as function of ε. At high values of divergence, the overlapping of the peaks leads to higher relative errors of ε with respect to what foreseen by the calculations in chapter III.

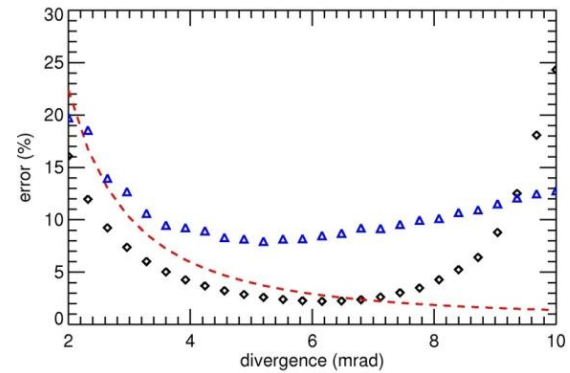

**Figure 5:** Accuracy of the measurements of ε made on BES spectra simulated for various values of ε. The diamonds and the triangles refer to the analysis of spectra simulated for vertical and horizontal LOSs respectively. The dashed line instead indicates the relative error foreseen by the preliminary calculations described in chapter III. As in Figure 1, the results are calculated for D operations.

## VII. CONCLUSIONS

The design of BES diagnostic for the MITICA experiment was optimized in order to minimize the measurement error of the beamlet divergence. A comprehensive numerical model was then set up to reproduce the spectra obtained with the diagnostic. The analyses of these spectra have confirmed that the diagnostic will be able to measure values of divergence between 3 mrad and 7 mrad e-folding with an error within or slightly higher than 10%.

## ACKNOWLEDGEMENTS

This work was set up in collaboration and financial support of Fusion for energy.